\documentclass[reprint,aps,prb,english,superscriptaddress,floatfix]{revtex4-2}

\usepackage{graphicx}
\usepackage{bm,amssymb,amsmath,mathrsfs,amstext,latexsym,physics}
\usepackage[T1]{fontenc}

\usepackage[usenames,dvipsnames]{xcolor}
\usepackage[colorlinks=true,citecolor=blue,urlcolor=blue,linkcolor=blue]{hyperref}
\usepackage[normalem]{ulem}

\usepackage{xcolor}

\usepackage{soul}
\setstcolor{red}

\def\rmd{\mathrm{d}}

\begin{document}

\title{Two-dimensional stealthy hyperuniform polycrystalline disk packings}

\author{Carlo Vanoni}
\affiliation{Department of Physics, Princeton University, Princeton, New Jersey 08544, USA}

\author{Paul J. Steinhardt}
\affiliation{Department of Physics, Princeton University, Princeton, New Jersey 08544, USA}

\author{Salvatore Torquato}
\affiliation{Department of Chemistry, Princeton University, Princeton, New Jersey 08544, USA}
\affiliation{Department of Physics, Princeton University, Princeton, New Jersey 08544, USA}
\affiliation{Princeton Institute for the Science and Technology of Materials, Princeton University, Princeton, New Jersey 08544, USA}
\affiliation{Program in Applied and Computational Mathematics, Princeton University, Princeton, New Jersey 08544, USA}

\begin{abstract}
    Polycrystals consist of grains of local crystalline order separated by grain boundaries. Conventional polycrystals are generally not hyperuniform---unlike perfect crystals, they lack anomalously suppressed large-scale density fluctuations---because their grains are randomly sized and oriented.
    In this paper, we use a collective-coordinate optimization procedure to generate two-dimensional polycrystalline packings composed of identical disks arranged in a pattern that is ultradense and stealthy hyperuniform (SHU). Stealthiness further strongly suppresses density fluctuations over a finite range of long wavelengths around the origin.  
    We compare them with polycrystalline disk packings obtained via a modified Lubachevsky--Stillinger (LS) rapid compression algorithm, a molecular dynamics protocol that serves as a standard reference model describing realistic, nonhyperuniform polycrystalline microstructures.
    We carry out an extensive comparison of polycrystalline SHU and LS packings that includes differences in two-point statistics, grain size, specific surface area, diffusion spreadability, and optical response as quantified by the imaginary part of the effective dynamic dielectric constant. 
    We find that the polycrystalline SHU packings exhibit a distinctive grain-size distribution, a consequence of long-range correlations between different grains that is absent in the nonhyperuniform case. 
    Within the nonlocal strong-contrast expansion, we confirm that polycrystalline SHU packings made of dielectric material are perfectly transparent to electromagnetic waves at small wave vectors, in contrast to LS packings. Moreover, polycrystalline SHU packings offer enhanced diffusion spreadability.
    Although polycrystalline SHU packings are not generally expected to form spontaneously, long-range interactions or correlated substrates may provide possible routes to their natural formation. Nonetheless, it is noteworthy that they can be created as metamaterials via nanolithography or 3D printing.
\end{abstract}

\maketitle

\section{Introduction}
\label{sec:Intro}

Polycrystalline materials consist of finite crystalline grains separated by grain boundaries. Thus, local order coexists with large-scale disorder. This grain-scale disorder degrades scattering and transport but improves ductility, fracture control, and strain hardening~\cite{As76,Ki05,Chaik95,Sutton2020Elasticity}. Typical experimental and simulated polycrystals are therefore expected to be nonhyperuniform: finite grains broaden Bragg peaks~\cite{Scherrer1918,Patterson1939Scherrer}, while grain-scale heterogeneity sustains long-wavelength density fluctuations, so that the structure factor $S(\mathbf{k})$ does not vanish as the wave number $k\equiv |\mathbf{k}|$ goes to zero~\cite{Hull2011Dislocations,X-rayScattering}.
These physical structures can be modeled using polycrystalline hard disk packings in which the disk centers form locally crystalline domains separated by grain boundaries. In these packings, the disk centers define a point configuration whose structure factor directly probes the suppression or persistence of large-scale density fluctuations. 

\begin{figure*}
    \centering
    \includegraphics[width=\textwidth]{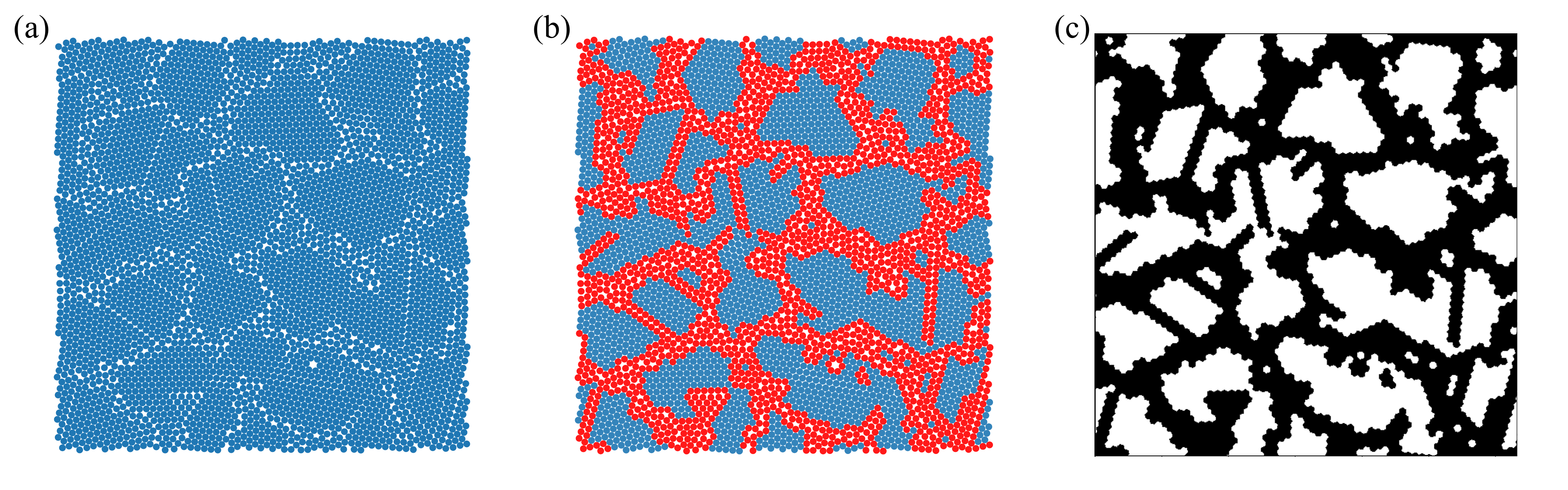}
    \caption{(a) Configuration of $N=4000$ disks obtained with the collective coordinate optimization procedure for $\chi = 0.0025$ and $\phi = 0.86$. This configuration can be interpreted as a two-phase medium, with disks constituting one phase, and the space between them forming the second phase. (b) Each disk is assigned to a grain (blue) or grain boundary (red) depending on whether or not it has six neighbors within a distance $d < 2 a(1 + \tau)$, where $a = \sqrt{\phi/\pi}$ is the disk radius at unit density and $\tau = 0.08$ (see text for details). (c) We can then construct another two-phase medium in which the set of grains constitutes one phase (in white), and the boundaries another phase (in black).}
    \label{fig:point-to-2phase}
\end{figure*}

In this paper, we show that it is possible to construct a heretofore unknown type of polycrystalline packings of identical circular disks that, despite their defects (e.g., grain boundaries), are not only hyperuniform~\cite{To03a,To18a}, meaning that they strongly suppress density fluctuations at the largest length scales so that $S(\mathbf{k})\to 0$ in the limit $k\to 0$, but ``stealthy'' (as defined below).
Our approach is to numerically construct these disk packings beginning from ultradense stealthy hyperuniform (SHU) point patterns~\cite{Kim_2025_DenseSphere} (see~\autoref{fig:point-to-2phase} for an example). The disks comprising the individual grains are arranged with hexagonal crystalline order and, even though they are separated by a complex disordered-looking network of grain boundaries, the entire configuration---grains and grain boundaries together---satisfies the stealthy hyperuniform condition.
These stealthy hyperuniform polycrystalline structures are not generally expected to form spontaneously. However, correlated substrates near commensurate--incommensurate transitions~\cite{Coppersmith1982}, or long-range interactions in charged dusty plasmas, could provide mechanisms for imposing the required correlations. They also represent templates for constructing metamaterials that combine properties of polycrystals and disordered stealthy hyperuniform solids.

Within the broader class of hyperuniform systems, a special subclass is given by stealthy hyperuniform configurations~\cite{Uc04b,batten_classical_2008,torquato_ensemble_2015,Zh15a}. These are defined by a structure factor that vanishes over a finite range of wave numbers around the origin,
\begin{equation}
\label{eq:Sk_SHU}
    S(\mathbf{k}) = 0 \quad \mathrm{for} \quad 0 < k \leq K .
\end{equation}
SHU systems may be either ordered—such as perfect crystals—or disordered~\cite{Uc04b,torquato_ensemble_2015,Zh15a}.
Stealthy configurations are further characterized by the stealthiness parameter $\chi$ [see Eq.~\eqref{eq:chi1}], which quantifies the fraction of independently constrained wave vectors relative to the total number of degrees of freedom~\cite{batten_classical_2008}. Disordered SHU systems combine liquid-like statistical isotropy with crystal-like suppression of long-wavelength scattering. This combination has been shown to produce unusual transport~\cite{To21d} and wave-propagation properties, including transparency regimes~\cite{Kim_2024_extraordinary}, and modified localization behavior~\cite{vanoni2025effective}. 

The consequences of local defects in “perfect'' hyperuniform systems have been studied by Kim and Torquato in Ref.~\cite{Ki18a}. They quantified how uncorrelated point defects, stochastic particle displacements, and thermal excitations generically degrade hyperuniformity. 
However, the SHU polycrystal disk packings introduced in this paper are exceptional because they do not follow the usual degradation of hyperuniformity despite the variation in grain sizes and orientations and the complex network of grain boundaries. 

We construct the polycrystalline disk packings numerically by first using a collective coordinate optimization procedure (CCO)~\cite{Kim_2025_DenseSphere,Torquato2025Existence} to generate ultradense 2D SHU point configurations subject to a soft-core repulsive potential; then we decorate each point with a hard disk of identical size such that there is no overlap. In particular, in the small-$\chi$ limit (we will fix $\chi = 0.0025$ from now on), the resulting configurations are not jammed and contain grain boundaries, yet allow us to obtain polycrystalline disk packings in the range $\phi \in [0.80,0.86]$. (The configurations do not display crystalline grains for $\phi \lesssim 0.80$.) 

To benchmark the characteristics of the SHU polycrystals, we also generate disk packings using a modified version of the Lubachevsky--Stillinger (LS) molecular-dynamics algorithm~\cite{Lu90,Do05a,Sk06}, described in~\autoref{sec:LSalgorithm}. LS polycrystals commonly serve as a reference model for polycrystalline microstructures of real materials, capturing the grain-size and orientation disorder that are typically observed.
By analogy with the case of jammed packings, we expect other rapid-compression protocols to produce qualitatively the same outcome~\cite{Do05d,Ma23,To10e,Ch12,Oh02,Wi21}.

After assigning each disk to a grain or grain boundary based on the number of close neighbors (see~\autoref{fig:point-to-2phase} and~\autoref{sec:grain_membership} for details), we perform a statistical comparison of the polycrystalline disk packings generated with the two methods, computing the structure factor, the grain-boundary surface distribution, and the grain-size distribution, and the spectral density and autocorrelation function of the two-phase medium formed by grains and boundaries. While some of these quantities do not show a qualitative difference between SHU and LS polycrystals, the grain-size distribution shows a significant difference between the two cases: the grain-size distribution of the SHU polycrystals displays a peak at a typical grain size for large packing fractions, which indicates long-range correlations between different grains, which is absent in the nonhyperuniform case.

We also compare the physical properties of the two types of polycrystals. In particular, we examine diffusive transport via the time-dependent diffusion spreadability~\cite{To21d}, and optical properties via the effective dynamic dielectric constant~\cite{To21a}. We find that, in the long-time limit, SHU polycrystals exhibit significantly larger spreadability than the LS polycrystals, in agreement with the spectral density being exactly zero for $0\leq k \leq K$. In addition, within the exact nonlocal strong-contrast expansions recently derived by Torquato and Kim~\cite{To21a}, we find that SHU polycrystals display a small interval of perfect transparency for electromagnetic waves at long wavelengths. 
 These theoretical results indicate differences that may be advantageous for metamaterials based on SHU polycrystal patterns, demonstrating that hyperuniformity is not only useful in cases where local crystalline order and longer-range disorder can be combined in a way to maintain stealthy hyperuniformity overall.

The remainder of the paper is organized as follows.
In~\autoref{sec:Background}, we collect the theoretical background and technical definitions used throughout the paper, including hyperuniformity, two-phase media, and the nonlocal strong-contrast formalism.
In~\autoref{sec:polycrystalformation}, we describe the two classes of polycrystalline packings studied in this work: SHU packings generated by collective-coordinate optimization and LS packings generated by a modified molecular-dynamics compression protocol. We also introduce the grain-membership criterion used to distinguish crystalline grains from grain boundaries.
In~\autoref{sec:structural}, we characterize the structural properties of these polycrystals, including two-point statistics and grain-size (given by the number of disks in a grain) distributions.
In~\autoref{sec:physical}, we examine their physical response, focusing on diffusion spreadability and optical transparency.
Finally, in~\autoref{sec:conclusion}, we summarize our conclusions.

\section{Background and Methods}
\label{sec:Background}

In this section, we collect the theoretical definitions and formal results used throughout the paper. We first review hyperuniformity and stealthiness, then summarize the two-phase-medium formalism and the nonlocal strong-contrast expansion used to analyze optical response.

\subsection{Hyperuniformity}
\label{sec:hyperunif}

Consider $N$ points in $d$-dimensional Euclidean space $\mathbb{R}^d$ at number density $\rho$, and let $\Omega(R)$ denote a spherical observation window of radius $R$. If $N(R)$ is the number of points in $\Omega(R)$, then the corresponding number variance $\sigma_N^2(R) = \langle N(R)^2\rangle - \langle N(R)\rangle^2$ provides a convenient criterion for distinguishing hyperuniform systems from typical disordered systems through its large-$R$ scaling.

An equivalent characterization is given in reciprocal space through the structure factor $S(\mathbf{k})$, which is related to $\sigma_N^2(R)$ by~\cite{To03a,To18a}
\begin{equation}
\label{eq:sigma_F_space}
    \sigma^2_N(R)=\langle N(R)\rangle\left[\frac{1}{(2\pi)^d}\int_{\mathbb{R}^d} S(\mathbf{k})\,\tilde{\alpha}(k;R)\,\rmd\mathbf{k}\right],
\end{equation}
where $\tilde{\alpha}(k;R)$ is the Fourier transform of the scaled intersection volume of two spherical windows separated by a distance $r$ and, for spherical windows, is given explicitly by
\begin{equation}
    \tilde{\alpha}(k;R)=2^d\pi^{d/2}\Gamma(1+d/2)\,\frac{\left[J_{d/2}(kR)\right]^2}{k^d}.
\end{equation}
Equation~\eqref{eq:sigma_F_space} shows that the asymptotic growth of $\sigma_N^2(R)$ is controlled by the small-$k$ behavior of $S(\mathbf{k})$. In particular, if $S(\mathbf{k})\sim |\mathbf{k}|^\alpha$ as $|\mathbf{k}|\to 0$, then the corresponding large-$R$ scaling of $\sigma_N^2(R)$ falls into one of three hyperuniform classes~\cite{Za09,To18a}:
\begin{align}
    \sigma_N^2(R)\sim
    \begin{cases}
        R^{d-1}, & \alpha>1 \quad \text{(Class I)},\\
        R^{d-1}\ln R, & \alpha=1 \quad \text{(Class II)},\\
        R^{d-\alpha}, & 0<\alpha<1 \quad \text{(Class III)}.
    \end{cases}
\end{align}
Classes I and III are the strongest and weakest forms of hyperuniformity, respectively.

A wide variety of disordered hyperuniform systems are now known. Examples include perfect glasses~\cite{Zh16a}, one-component plasmas~\cite{Le00}, critical absorbing states in random organization models~\cite{He15,Wiese2024Hyperuniformity,Ma19}, maximally random jammed states~\cite{To00b,Ma23}, systems arising in random matrix theory and number theory~\cite{To08b,Mon73}, cold-atom processes~\cite{Le14}, active particle systems \cite{Le19b,backofen_nonequilibrium_2024}, biological systems~\cite{Ji14,Ma15,Hu21,ge_hidden_2023}, and multihyperuniform systems~\cite{Lo18a,christogeorgos2024computational}. See Ref.~\cite{To18a} for a review and additional references.

Among Class-I systems, stealthy hyperuniform (SHU) point patterns may be viewed as an extreme limit in which scattering is completely suppressed over a finite neighborhood of the origin: there exists a cutoff $K>0$ such that
\begin{equation}
\label{eq:stalthy_Sk}
    S(\mathbf{k})=0 \qquad \text{for}\quad 0<|\mathbf{k}|\le K.
\end{equation}
Crystals are trivial examples of stealthy hyperuniform systems, since their Bragg spectra imply $S(\mathbf{k})=0$ for all $|\mathbf{k}|$ smaller than the location of the first Bragg peak.

Disordered SHU materials, despite being statistically isotropic and lacking Bragg peaks, retain key crystal-like features, including the absence of single scattering from intermediate to infinite wavelengths [see Eq.~\eqref{eq:Sk_SHU}] and the prohibition of arbitrarily large holes in the thermodynamic limit~\cite{zhang_can_2017,ghosh_generalized_2017,To18a}. This hybrid crystal–liquid character endows disordered SHU materials with exceptional transport, elastic, and wave-propagation properties among isotropic amorphous states~\cite{To18a}. Notable examples include wave transparency~\cite{Le16,Fr17,To21a,Fr23,Kl22,kim_effective_2023,alhaitz_experimental_2023,Kim_2024_extraordinary}, enhanced wave absorption~\cite{Bi19}, improved solar-cell efficiency~\cite{merkel_stealthy_2023}, tunable localization and diffusion regimes~\cite{Fr17,Sg22,Sc22}, distinctive phononic responses~\cite{Gk17,Ro19,Roh20}, Luneburg lenses with reduced backscattering~\cite{Zh19}, extraordinary phased arrays~\cite{Ch21,tang2023hyperuniform}, optimal sampling arrays for 3D ultrasound imaging~\cite{tamraoui_hyperuniform_2023}, high-quality optical cavities~\cite{granchi_nearfield_2023}, network materials with nearly maximal effective electrical conductivities and elastic moduli~\cite{To18c}, particulate media with near-maximal effective diffusion coefficients~\cite{zhang_transport_2016}, and larger localization length in 1D disordered quantum systems, resulting in effective delocalization~\cite{vanoni2025effective}.

Disordered SHU systems exhibit additional geometric constraints, including the bounded-hole property. Here, a ``hole'' refers to a spherical region in $\mathbb{R}^d$ of a given radius that contains no point centers. It was conjectured~\cite{zhang_can_2017} and subsequently proven~\cite{ghosh_generalized_2017} that disordered SHU configurations cannot support arbitrarily large holes in the infinite-system-size limit, whereas typical disordered systems retain a nonzero probability of finding holes of arbitrarily large size in the thermodynamic limit.

The degree of stealthiness is conveniently quantified by the parameter $\chi$, which measures the fraction of constrained collective degrees of freedom. If $M(K)$ denotes the number of independently constrained wave vectors within the exclusion region $0<|\mathbf{k}|\le K$, then
\begin{equation}
\label{eq:chi1}
    \chi=\frac{M(K)}{d(N-1)},
\end{equation}
where $d(N-1)$ is the number of configurational degrees of freedom after removing translations. In the thermodynamic limit, one has~\cite{torquato_ensemble_2015}
\begin{equation}
\label{eq:stealthy_per}
    \chi=\frac{v_1(K)}{2\,\rho\,d\,(2\pi)^d},\qquad
    v_1(R)=\frac{\pi^{d/2}}{\Gamma(1+d/2)}R^d,
\end{equation}
with $v_1(R)$ the volume of a $d$-dimensional sphere of radius $R$. 

As discussed in~\autoref{sec:CollCoord}, SHU configurations can be constructed as ground states of an interaction potential designed to enforce the constraints in Eq.~\eqref{eq:stalthy_Sk}. For $\chi\gtrsim 0$, the ground-state manifold is highly degenerate, and the resulting patterns are markedly disordered, whereas increasing $\chi$ progressively reduces the dimensionality of the accessible configuration space. In particular, for $0\le \chi\le 1/2$, the dimensionality per particle satisfies $d_C=d(1-2\chi)$~\cite{torquato_ensemble_2015}, vanishing at $\chi=1/2$, where crystallization can occur depending on $d$. Periodic ground states exist throughout, but they form a zero-measure subset and are therefore not typically accessed by the collective-coordinate optimization procedure.

\subsection{Two-phase media}
\label{sec:TwoPhaseMedia}

We briefly introduce the two-phase media formalism that will be useful in the next sections.
Consider a medium composed of two phases, $i=1,2$. In this paper, we consider two distinct two-phase representations. In one, phase $i=1$ is the matrix filling the space between disks, and phase $i=2$ is the disk phase; in the other, phase $i=1$ is the polycrystalline-grain phase, and phase $i=2$ is the grain-boundary phase. The definitions below, however, apply to general two-phase media.

We define the indicator function
\begin{equation}
\mathcal{I}^{(i)}(\mathbf{x}) =
\begin{cases}
1 & \mathrm{if}\ \mathbf{x}\in \mathcal{V}_i,\\
0 & \mathrm{otherwise},
\end{cases}
\end{equation}
where $\mathcal{V}_i$ denotes the region occupied by phase $i$ with volume fraction $\phi_i$~\cite{To02a}.
The $n$-point correlation function of phase $i$ is
\begin{equation}
S_n^{(i)} = \left\langle \prod_{j=1}^{n} \mathcal{I}^{(i)}(\mathbf{x}_j) \right\rangle ,
\end{equation}
where the average is taken over realizations. For statistically homogeneous media $S_1^{(i)}=\phi_i$.
The autocovariance function is defined as
\begin{equation}
\label{eq:autocov}
\chi_V(\mathbf{r}) \equiv S_2^{(1)}(\mathbf{r})-\phi_1^2
= S_2^{(2)}(\mathbf{r})-\phi_2^2 ,
\end{equation}
which is identical for both phases and satisfies $\chi_V(\mathbf{0})=\phi_1\phi_2$.
Its Fourier transform defines the spectral density
\begin{equation}
\tilde{\chi}_V(\mathbf{k}) =
\int_{\mathbb{R}^d} d\mathbf{r}\,
\chi_V(\mathbf{r}) e^{-i\mathbf{k}\cdot\mathbf{r}},
\end{equation}
with $\tilde{\chi}_V(\mathbf{k})\ge0$.

In the specific case where packings of identical spheres of radius $a$ form one of the two phases, the spectral density depends only on $k=|\mathbf{k}|$ and is related to the structure factor $S(k)$~\cite{To02a,To16b} by
\begin{equation}
\label{eq:spectr_dens_1}
\tilde{\chi}_V(k)=\phi_2\,\tilde{\alpha}_2(k;a)\,S(k),
\end{equation}
where
\begin{equation}
\label{eq:alpha_tilde}
\tilde{\alpha}_2(k;a)=
\frac{1}{v_1(a)}
\left(\frac{2\pi a}{k}\right)^d
J_{d/2}^2(ka).
\end{equation}
Here, $v_1(a)=\pi^{d/2}a^d/\Gamma(1+d/2)$ is the $d$-dimensional sphere volume and $\phi_2=\rho v_1(a)$.
For stealthy hyperuniform packings, $S(k)=0$ for $0<k\le K$, implying $\tilde{\chi}_V(k)=0$ in the same interval.
At large $k$, the spectral density scales as $\tilde{\chi}_V(k)\sim \gamma(d)s/k^{d+1}$, where $\gamma(d)$ is a dimension-dependent constant and $s$ is the specific surface~\cite{To21d}.

\subsection{Nonlocal strong contrast expansion}
\label{sec:strongcontrast_background}

Full derivations and additional details can be found in Refs.~\cite{To21a,Kim_2024_extraordinary}. We recall that the expansion is not performed directly on the effective dielectric tensor $\epsilon_e(\mathbf{k}_q)$, but rather on the transformed quantity $\left(L_e^{(q)}(\mathbf{k}_q)\right)^{-1}$, defined as
\begin{equation}
    L_e^{(q)}(\mathbf{k}_q) \equiv \left(\epsilon_e(\mathbf{k}_q) - \epsilon_q\mathbf{I}\right)
\cdot \left[ \mathbf{I} + \mathbf{D}^{(q)}
\cdot \left(\epsilon_e(\mathbf{k}_q) - \epsilon_q\mathbf{I}\right) \right]^{-1}.
\end{equation}
Here $p$ and $q$ ($p\ne q$) label the two phases, with $q$ denoting the reference phase. The nonlocal strong-contrast expansion then takes the form
\begin{equation}
\label{eq:NLSCE}
    \phi_p L_p^{(q)} \cdot \left(L_e^{(q)}(\mathbf{k}_q)\right)^{-1} \cdot \phi_p L_p^{(q)}
= \phi_p L_p^{(q)} - \sum_{n=2}^\infty \mathcal{A}_n^{(p)}(\mathbf{k}_q),
\end{equation}
where we have defined
\begin{equation}
    L_p^{(q)} \equiv (\epsilon_p - \epsilon_q)
\left[ \mathbf{I} + \mathbf{D}^{(q)} (\epsilon_p - \epsilon_q) \right]^{-1}.
\end{equation}
For our purposes, we consider 3D media whose transverse sections correspond to 2D disk packings. For a normally incident wave with $\mathbf{k}_q = k_q \hat{\mathbf{y}}$, symmetry implies that the effective dielectric tensor decomposes into transverse magnetic (TM) and transverse electric (TE) components.

Truncating the expansion at second order yields explicit expressions for $\epsilon_e^{TM}(k_q)$ and $\epsilon_e^{TE}(k_q)$ in terms of phase contrast, volume fraction, and microstructural information (spectral density)~\cite{Kim_2024_extraordinary}:
\begin{widetext}
\begin{equation}
\label{eq:TM_2D}
    \frac{\epsilon_e^{TM}(k_q)}{\epsilon_q} =
    1 + \frac{\phi_p^2 \left[(\epsilon_p + \epsilon_q)\beta_{pq}\right]/\epsilon_q}
    {\phi_p - A_2^{TM}(k_\ast^{TM}, \langle \epsilon \rangle) \left[(\epsilon_p + \epsilon_q)\beta_{pq}\right]},
\end{equation}

\begin{equation}
\label{eq:TE_2D}
    \frac{\epsilon_e^{TE}(k_q)}{\epsilon_q} =
    1 + \frac{2\phi_p^2\beta_{pq}}
    {\phi_p(1 - \phi_p\beta_{pq})
    - A_2^{TE}\left(k_\ast^{TE}; \epsilon_{BG}^{(2D)}\right)\left[2\epsilon_q\beta_{pq}\right]},
\end{equation}
where $\langle\epsilon\rangle$ is the arithmetic mean of the local dielectric constant, $\beta_{pq} = (\epsilon_p - \epsilon_q)/(\epsilon_p + (d-1)\epsilon_q)$ (we set $d=2$), $k_*^{TE} \equiv k_q \sqrt{\epsilon_{BG}^{(2D)}/\epsilon_q}$, $k_*^{TM} \equiv k_q \sqrt{\langle \epsilon \rangle/\epsilon_q}$, and $\epsilon_{BG}^{(2D)}$ is the Bruggeman approximation for two-phase media~\cite{To02a,bruggeman_berechnung_1935}.
The second-order coefficients $A_2^{TM}(k_q; \epsilon_q)$ and $A_2^{TE}(k_q; \epsilon_q)$ are given by
\begin{align*}
\label{eq:A_2D}
    A_2^{TM}(k_q; \epsilon_q) &= 2 A_2^{TE}(k_q; \epsilon_q) = -\frac{\pi}{2\epsilon_q}F^{(2D)}(k_q) \\
    &= \frac{1}{\epsilon_q} \left\{ \frac{k_q^2}{\pi^2} \int_0^{\pi/2} d\phi
    \left[ \text{p.v.} \int_0^{\infty} dq \frac{2q \tilde{\chi}_V(q)}{q^2 - (2k_q \cos\phi)^2} \right]
    + i \frac{k_q^2}{\pi} \int_0^{\pi/2} \tilde{\chi}_V(2k_q \cos\phi) d\phi \right\}.
\end{align*}
\end{widetext}
For the disk--matrix calculations, $p=2$ denotes the disk phase and $q=1$ denotes the matrix, consistent with the phase definitions in Sec.~\ref{sec:TwoPhaseMedia}. The advantage of the expansion in Eq.~\eqref{eq:NLSCE}, formulated as a linear fractional transformation of $\epsilon_e(\mathbf{k}_q)$, lies in its rapid convergence when truncated at low order on the right-hand side. In practice, Eqs.~\eqref{eq:TM_2D} and~\eqref{eq:TE_2D} are accurate approximations to the full series, as they effectively incorporate higher-order contributions through lower-order diagrammatic terms~\cite{To21a}.

\section{Polycrystalline packings}
\label{sec:polycrystalformation}

In this section, we generate two different ensembles of polycrystal packings and discuss the procedure used to identify grains and grain boundaries. The first ensemble consists of ultradense SHU disk packings generated by collective-coordinate optimization with an additional short-range repulsive constraint. The second consists of nonhyperuniform polycrystalline disk packings generated by a modified Lubachevsky--Stillinger molecular-dynamics protocol. We then define a grain-membership criterion tailored to the nonjammed polycrystalline packings considered here.

\subsection{Collective coordinate optimization and SHU polycrystalline packings}
\label{sec:CollCoord}

We generate disordered SHU configurations using a modified collective coordinate optimization (CCO) scheme~\cite{torquato_ensemble_2015}. Following Ref.~\cite{zhang_can_2017}, we construct an objective function whose ground states simultaneously enforce the stealthy constraints in reciprocal space and a minimum pair separation in real space. Specifically, we minimize the potential energy
\begin{equation}
\label{eq:soft_core}
    \Phi(\mathbf{r}^N)=\frac{N}{2V_{\mathcal{F}}}\sum_{\mathbf{k}}\tilde{v}(\mathbf{k})\,S(\mathbf{k})+\sum_{i<j}u(r_{ij}),
\end{equation}
where $S(\mathbf{k})$ is the single-configuration structure factor, $V_{\mathcal{F}}$ is the real-space volume associated with the constrained region, and $\tilde{v}(\mathbf{k})=V(k)\,\Theta(K-k)$ with $\Theta$ is the Heaviside step function. Choosing a non-negative weight $V(k)\ge 0$ (e.g., $V(k)=1$) guarantees $\Phi(\mathbf{r}^N)\ge 0$, and $\Phi(\mathbf{r}^N)=0$ if and only if the stealthy condition $S(\mathbf{k})=0$ holds for all wave vectors with $0<|\mathbf{k}|\le K$.

To prevent disk overlap and control the packing fraction, we supplement the CCO objective with a short-range repulsive pair interaction $u(r)$ that penalizes separations smaller than a prescribed distance $\sigma$. In this work, we adopt a soft-core repulsion of the form
\begin{equation}
    u(r)=u_0\left(1-\frac{r}{\sigma}\right)^2\Theta(\sigma-r),
\end{equation}
which is continuous and differentiable at $r=\sigma$ and matches the choice in Refs.~\cite{Kim_2025_DenseSphere,Vanoni2025Dynamical}. With this addition, the ground states of $\Phi$ correspond to SHU configurations at cutoff $K$ with minimum pair distance $\sigma$. Although other choices of $u(r)$ do not change the qualitative nature of the ground-state ensemble, they can affect numerical efficiency. Given a SHU point configuration, one can then obtain a disk packing by decorating each point with a disk of radius $a=\sigma/2$.

The repulsive term introduces an additional constraint on the ground-state manifold, implying that not all $\sigma$ are admissible for fixed $(d,\chi,N)$. For a given $\chi$, increasing $\sigma$ reduces the dimensionality of the ground-state manifold and may drive it to zero at a critical value $\sigma=\sigma_c$~\cite{Kim_2025_DenseSphere}. The dependence of $\sigma_c$ on $\chi$ and $d$ has been studied in Ref.~\cite{Kim_2025_DenseSphere}; notably, $\sigma_c$ decreases with increasing $\chi$. In particular, the maximum packing fraction achievable using this procedure depends on the stealthiness $\chi$, reaching up to $\phi_{\mathrm{max}}(\chi, d) = 1.0$, $0.86$, $0.63$ in the zero-$\chi$ limit and decreasing to $\phi_{\mathrm{max}}(\chi, d) = 1.0$, $0.67$, $0.47$ at $\chi = 0.45$ for $d = 1$, $2$, $3$, respectively~\cite{Kim_2025_DenseSphere,Note1}. This trend reflects the fact that reducing $\chi$ relaxes the number of constrained stealthy modes and thereby allows more effective near-contact constraints, which in turn reduce void space and raise the maximal packing fraction.

We restrict the present comparison to $\chi=0.0025$ because the ultradense polycrystalline regime requires small $\chi$. Increasing $\chi$ changes both the stealthy cutoff and the maximum attainable packing fraction, while nonstealthy hyperuniform classes would require a different collective-coordinate target. Fixing $\chi$ therefore isolates the comparison between ultradense SHU and nonhyperuniform LS polycrystals at the same packing fraction.

In the limit $\chi \to 0$, ultradense SHU packings obtained through the CCO algorithms in two and three dimensions become configurationally nearly indistinguishable from the corresponding jammed hard-particle packings produced by rapid compression algorithms~\cite{Torquato2025Existence,Kim_2025_DenseSphere}, such as rapid compression of hard spheres~\cite{Do05d,Ma23,To10e,Ch12}, rapid quenching of soft spheres at high temperature to find inherent structures at zero temperature~\cite{Ch12,Oh02}, or as a dynamical phase transition via a biased random organization protocol~\cite{Wi21}.

Numerically, for given $(d,\chi,N,\sigma)$, we start from a random ``high-temperature'' configuration at unit density $\rho=1$ and minimize $\Phi$ in Eq.~\eqref{eq:soft_core} using the low-storage Broyden--Fletcher--Goldfarb--Shanno (L-BFGS) algorithm~\cite{Liu89}. We terminate the minimization once $\Phi<7\times 10^{-20}$. The initial configuration contains no prescribed grains or grain orientations; the grains emerge during minimization as a consequence of the simultaneous stealthy and minimum-separation constraints. Figure~\ref{fig:point-to-2phase} illustrates the subsequent conversion of the optimized point pattern into a disk packing and its grain--boundary representation. The statistical properties of the resulting ground states depend on both the initial condition and the choice of $V(k)$~\cite{Zh15a,Dawley2025Evolution}; for example, although periodic configurations belong to the ground-state manifold, they are entropically suppressed and are not typically reached from random initial conditions.

\subsection{Lubachevsky--Stillinger polycrystalline packings}
\label{sec:LSalgorithm}

A variety of numerical protocols have been developed to generate dense and jammed sphere packings, including event-driven growth algorithms, energy-minimization methods, and optimization-based compression schemes~\cite{Lu90,OH03,To09c,To10e,At14}.
In this paper, we benchmark the SHU polycrystalline packings with disk-packing configurations generated using a modified event-driven molecular-dynamics Lubachevsky--Stillinger (LS) algorithm~\cite{Lu90,Do05a}, which yields effectively collectively jammed hard-disk packings~\cite{Do04a} (nearly strictly jammed under fixed periodic boundary conditions, without deforming the fundamental cell), able to reach up to $\phi \simeq 0.88$. The resulting configurations, as a consequence of the absence of geometric frustration in 2D, consist of large triangular-lattice grains~\cite{Do04a,Torquato2026Review} and therefore provide a natural point of comparison for our results. 

In the LS algorithm, initially, Poisson-distributed points are grown into nonoverlapping disks of prescribed radius at a prescribed rate $\epsilon$. During the growth process, disk positions evolve according to Newtonian dynamics, with an additional kinetic-energy boost after collisions due to the positive expansion rate.  
This algorithm has been used to study the formation of amorphous jammed packings~\cite{Sk06} across spatial dimensions. That objective is achieved by choosing a large growth rate $\epsilon$ in the early stages, so that crystallization is avoided. In this paper, by contrast, we are interested in forming crystalline structures, and therefore we keep $\epsilon$ small even at short times~\cite{Do04a}.
As a consequence, the final state is generally not strictly jammed. This observation is important for defining the criterion used to assign each disk either to a grain of the polycrystal or to a grain boundary.

Representative SHU and LS configurations at $\phi=0.86$ are compared directly in Appendix~\ref{app:configurations}.

\subsection{Grain-membership criterion}
\label{sec:grain_membership}

\begin{figure}
    \centering
    \includegraphics[width=\linewidth]{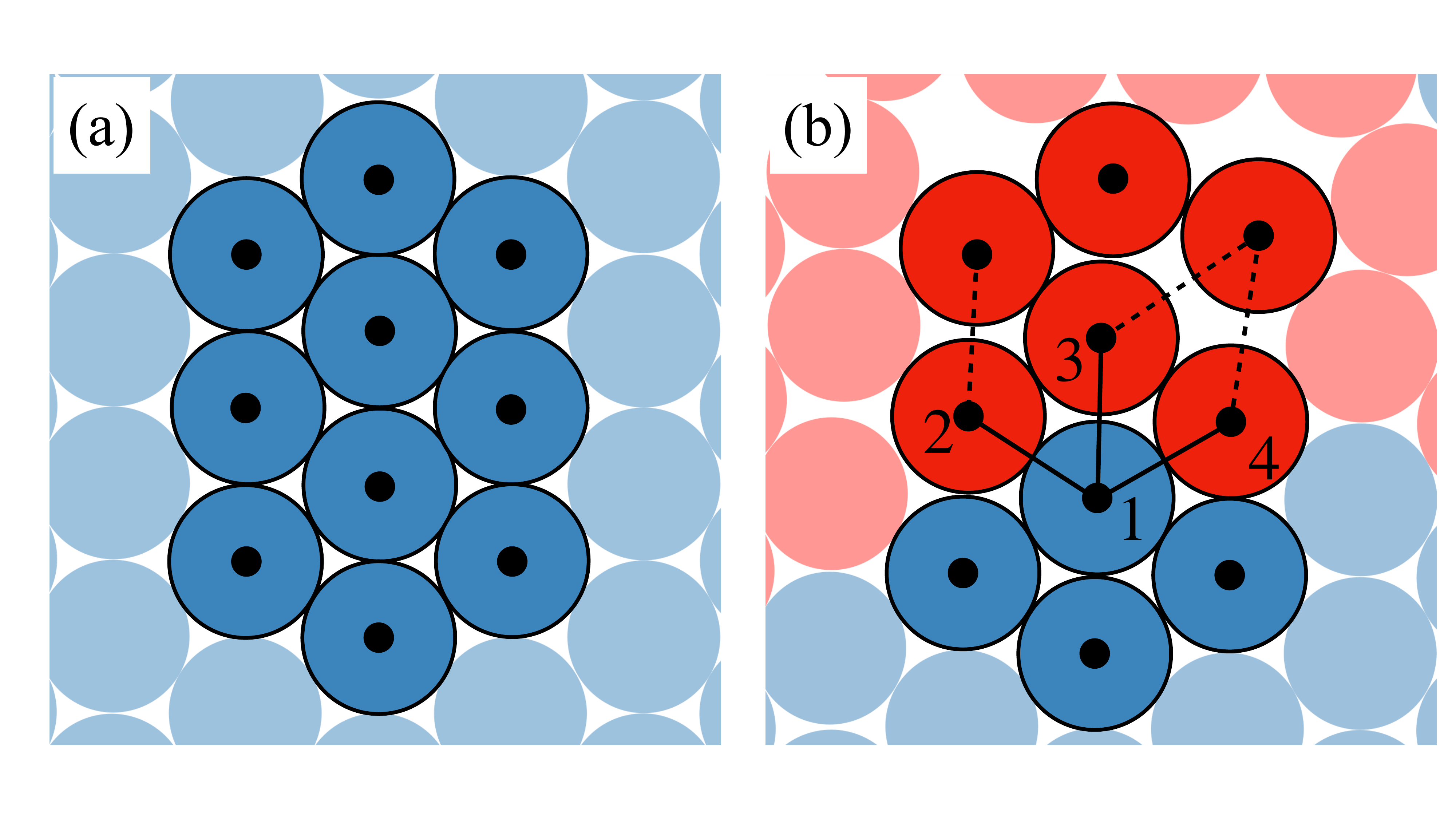}
    \caption{(a) If all disks have six neighbors at distance $d<d_{\tau}$, with $d_{\tau}$ defined in Eq.~\eqref{eq:dtau}, they are assigned to a grain (blue). (b) Near a grain boundary, some disks satisfy this condition (disk 1, solid lines), while others do not and are assigned to the boundary (disks 2, 3, and 4, dashed lines).}
    \label{fig:grain_membership}
\end{figure}

\begin{figure*}
    \centering
    \includegraphics[width=\linewidth]{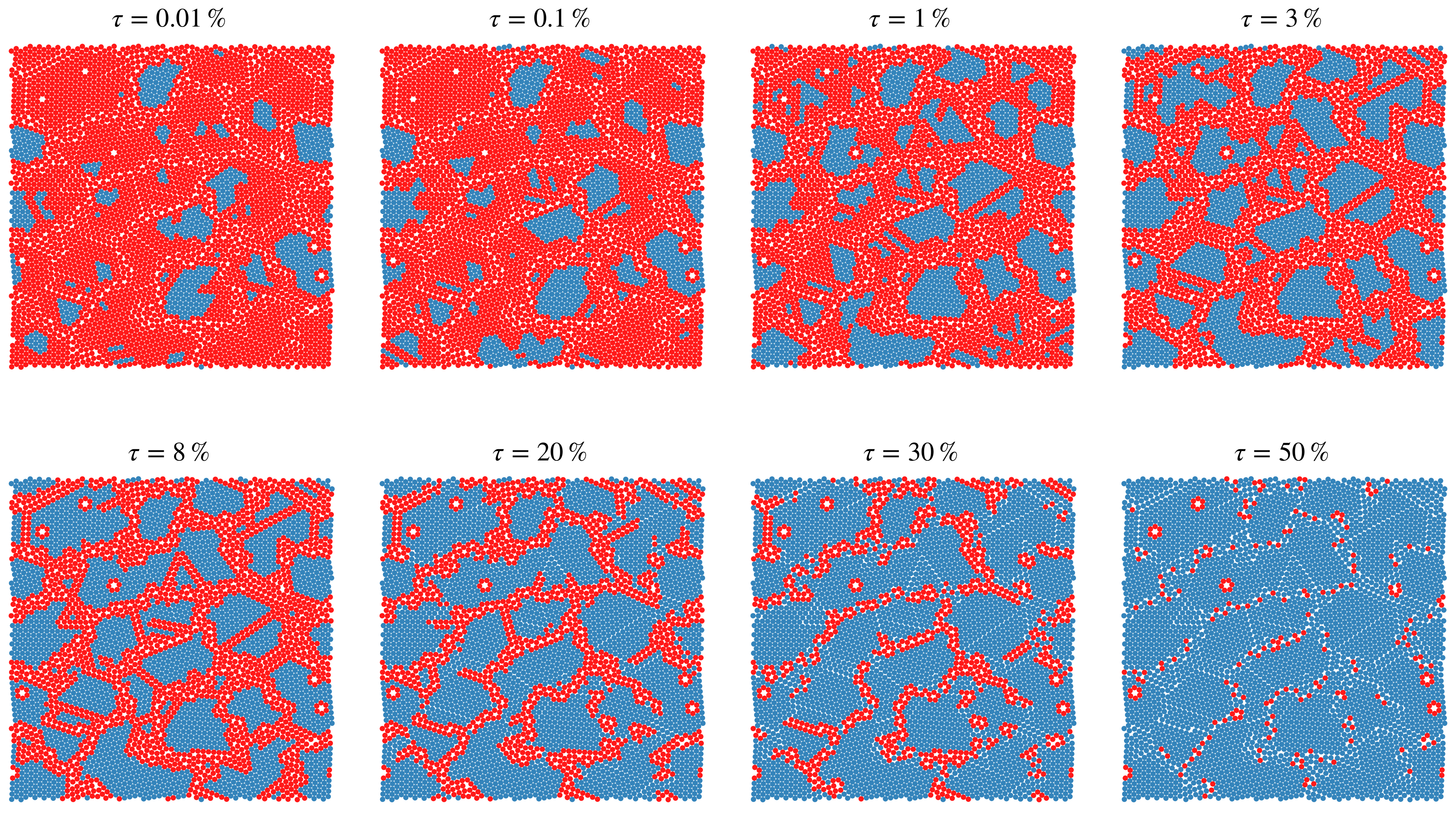}
    \caption{Grains (blue) and grain boundaries (red) for the same packing at different tolerance values $\tau$ [Eq.~\eqref{eq:dtau}]. For small $\tau$ (upper left), no grains are identified. 
    As $\tau$ increases, more disks are assigned to grains. For large $\tau$, almost all disks are assigned to grains. We use $\tau=0.08$, where the grain--boundary interface length has plateaued.}
    \label{fig:membership}
\end{figure*}

To structurally characterize polycrystals, we first define a grain-membership rule. Because crystalline grains are locally triangular, a disk is classified as belonging to a grain if it has six neighbors within a cutoff distance $d_{\tau}$, with $d_{\tau}\gtrsim 2a$ and $a=\sqrt{\phi/\pi}$ the disk radius (see~\autoref{fig:grain_membership}).
To be concrete, we define $d_{\tau}$ as
\begin{equation}
    d_{\tau} = 2a(1+\tau).
    \label{eq:dtau}
\end{equation}
Different choices of the tolerance $\tau$ yield different partitions into grains and grain boundaries, as shown in~\autoref{fig:membership}. For very small $\tau$, almost all disks are assigned to grain boundaries, whereas for large $\tau$, only a few disks are identified as outside grains.

In jammed systems, a very small tolerance $\tau=O(10^{-9})$, determined from the plateau of the cumulative coordination-number distribution versus gap tolerance~\cite{Do05c}, is sufficient to account for numerical precision and to classify a disk as belonging to a grain or a boundary. Here, the packings are not jammed, so that criterion does not apply; in particular, no plateau is visible at very small tolerances, and $\tau$ must be selected differently.
\begin{figure}
    \centering
    \includegraphics[width=\linewidth]{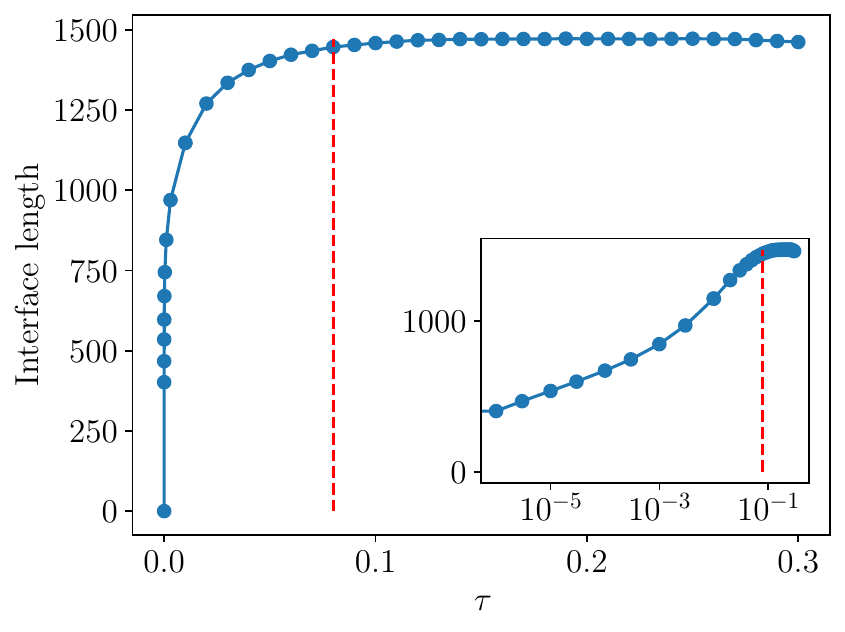}
    \caption{Grain--boundary interface length (in units of $\rho^{-1/2}$, with $\rho=1$), averaged over 100 SHU configurations at $\phi = 0.86$, as a function of the tolerance $\tau$. At small tolerance $\tau$, the average interface is near zero, consistent with the upper-left panel of~\autoref{fig:membership}. As $\tau$ increases, the interface increases and reaches a broad maximum plateau before returning toward zero at large $\tau$. We choose the optimal tolerance at the onset of this plateau (vertical red dashed line), namely, $\tau = 0.08$. The inset shows the same data on a log-linear scale.}
    \label{fig:specific_surface}
\end{figure}

We therefore use the total grain--boundary interface length as a function of $\tau$. This quantity is clearly positive and vanishes both at $\tau = 0$ and $\tau \gg 1$, and thus it is nonmonotonic and, as shown in~\autoref{fig:specific_surface}, it exhibits a broad plateau near its maximum at intermediate $\tau$. We choose the onset of this plateau $\tau=0.08$ (vertical red dashed line) as our working tolerance. Along the plateau, the mean grain size increases while the interface length remains nearly constant, indicating that progressively more interfacial disks are reclassified as grain disks (compare $\tau=8\%$ and $\tau=20\%$ in~\autoref{fig:membership}). Selecting the smallest plateau value, therefore, yields a conservative grain definition with fewer interfacial imperfections.

Once each disk is assigned either to a grain or to a grain boundary, we consider the Voronoi tessellation of the point process and assign each cell to a grain or boundary according to the classification of the corresponding disk.

This construction differs from an equilibrium polycrystal, whose boundary structure is selected by a physical free-energy functional. Our neighbor-count criterion identifies the crystalline grains and the noncrystalline interfacial regions. After identifying the grains, we estimate one representative orientation per grain from a nearest-neighbor bond direction, reduced modulo $\pi/3$; the resulting orientation distributions are reported in Appendix~\ref{app:grain_orientations}. The SHU distribution is broad and closely comparable to that of the LS polycrystals generated at the larger compression rate, $\epsilon=0.01$, showing that the SHU constraint is compatible with substantial orientational disorder among the grains. The LS distribution for $\epsilon=0.001$ is instead more concentrated around its circular mean because the same finite-size system contains fewer, larger grains. Here, $\tau$ controls the distance tolerance used to identify the interfacial regions. The orientation distribution characterizes the ensemble of grains, while a boundary-specific classification of twins or small- and large-angle boundaries would additionally require the pairwise misorientation of adjacent grains. Isolated disks assigned to the interfacial phase are treated as part of that phase rather than as equilibrium vacancies.

\section{Structural properties}
\label{sec:structural}

\subsection{Spectral density and autocorrelation function}
\label{sec:correlation}

We now determine the two-point correlation functions for two different two-phase media (see~\autoref{sec:TwoPhaseMedia} for more details on general properties of two-phase media). Throughout, lengths are expressed in units of $\rho^{-1/2}$, with $\rho=1$; correspondingly, $k$ and $\tilde{\chi}_V(k)$ are reported in units of $\rho^{1/2}$ and $\rho^{-1}$, respectively. First, consider the medium in which the matrix filling the space between disks is phase $1$, and the disks are phase $2$. In this case, Eq.~\eqref{eq:spectr_dens_1} applies, and the spectral density is proportional to the structure factor.
\begin{figure}
    \centering
    \includegraphics[width=\linewidth]{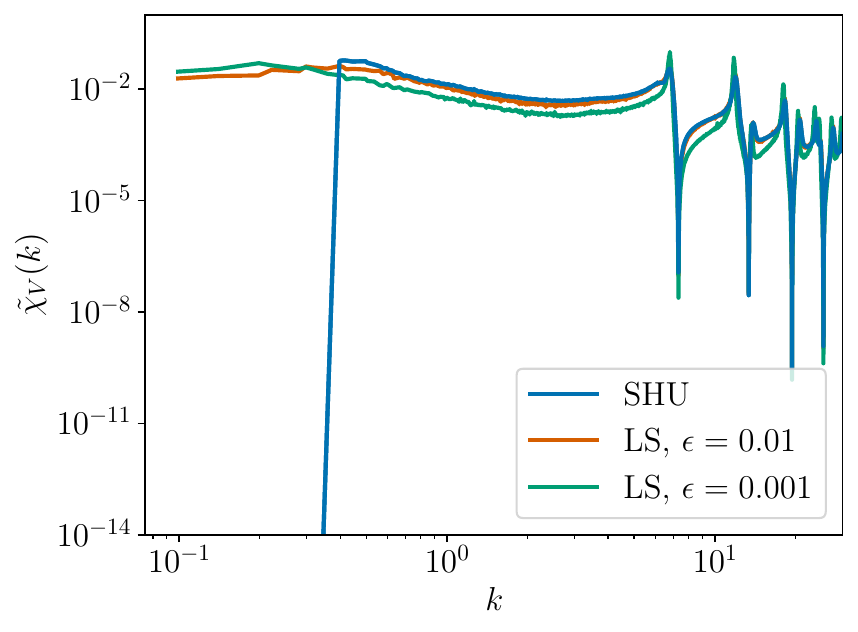}
    \caption{Spectral density $\tilde{\chi}_V(k)$ of the two-phase medium formed by disks and the matrix filling the space between them. SHU packings clearly display a stealthy region at small $k$, in which $S(k)$ and therefore $\tilde{\chi}_V(k)$ vanish, whereas LS packings retain finite spectral density down to $k=0$. SHU and LS packings with $\epsilon = 0.01$ show similar behavior at large $k$, while LS packings with $\epsilon=0.001$ exhibit more pronounced peaks, reflecting larger grains.}
    \label{fig:spectr_dens}
\end{figure}
As shown in Fig.~\ref{fig:spectr_dens}, the spectral density of media obtained from SHU disk packings vanishes for small $k<K = 4\sqrt{\pi \chi}$, in agreement with Eq.~\eqref{eq:stealthy_per}. By contrast, media obtained from LS packings have a finite spectral density that remains nonzero as $k\to 0$ and are therefore nonhyperuniform. In addition, two-phase media derived from SHU and LS packings with $\epsilon=0.01$ behave similarly at large $k$, whereas LS packings with $\epsilon=0.001$ show more pronounced peaks, consistent with their larger average grain size (see~\autoref{sec:grain_size}).

Second, we consider a two-phase medium in which grains constitute one phase and grain boundaries constitute the other [see~\autoref{fig:point-to-2phase}(c)]. As discussed in~\autoref{sec:grain_membership}, this medium is obtained from the Voronoi tessellation of the original disk packings, where each cell is assigned to a grain or boundary according to whether the corresponding disk belongs to a grain.
\begin{figure}[t!]
    \centering
    \includegraphics[width=\linewidth]{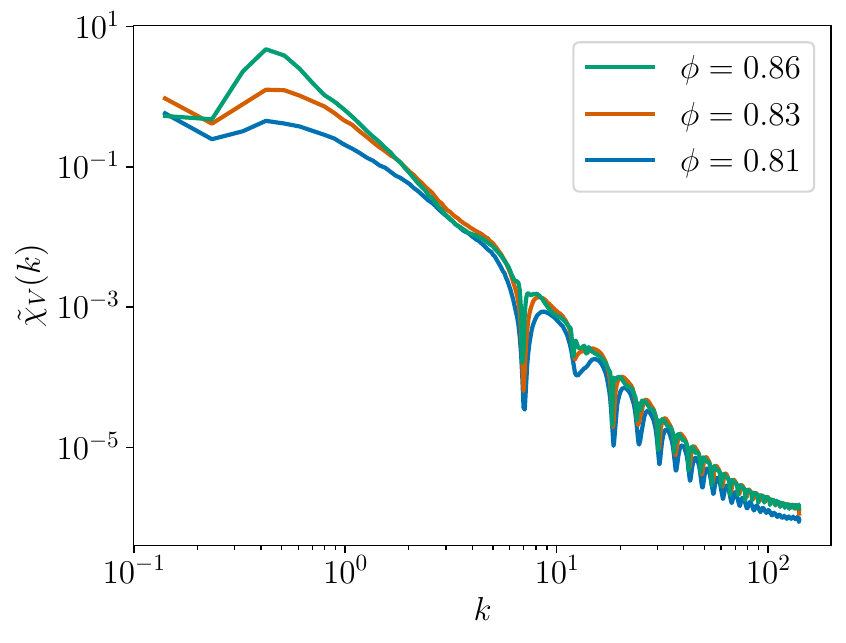}
    \caption{Spectral density of the two-phase medium formed by grains and grain boundaries obtained from SHU packings, for different packing fractions. As discussed in the main text, this medium is nonhyperuniform.}
    \label{fig:spectr_dens_grains_SHU}
\end{figure}
We report in~\autoref{fig:spectr_dens_grains_SHU} the results for media derived from SHU packings. 
As is clearly visible, stealthy hyperuniformity is lost under this transformation, and the resulting medium is nonhyperuniform. This is expected, since hyperuniformity is generally not preserved under such mappings.

\begin{figure}[t!]
    \centering
    \includegraphics[width=\linewidth]{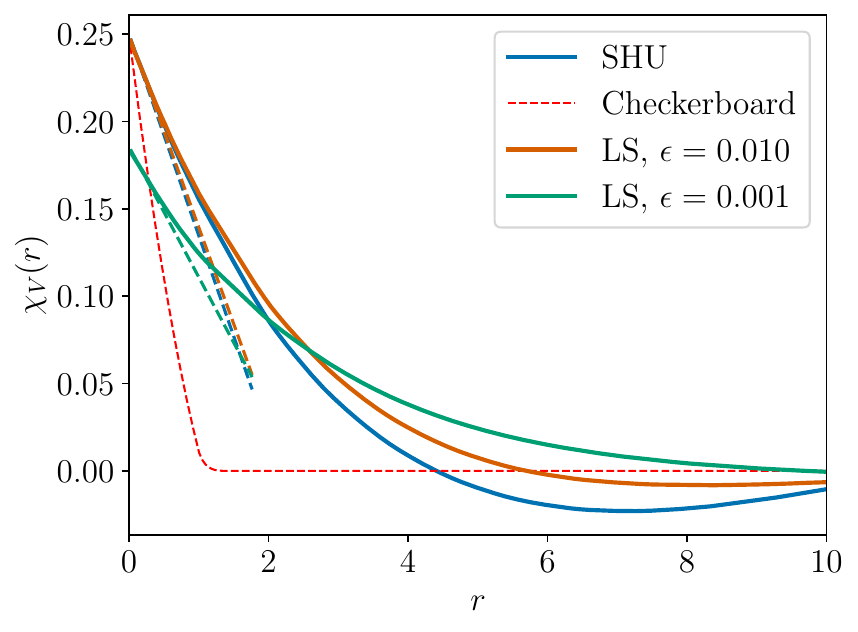}
    \caption{Autocorrelation function for the two-phase medium composed of grains and grain boundaries derived from disk packings. Dashed colored lines show the small-distance approximation in Eq.~\eqref{eq:autocorr_approx}. The red dashed line corresponds to the random-checkerboard two-phase medium (analytic expression in Ref.~\cite{To02a}).}
    \label{fig:autocorr}
\end{figure}

The autocorrelation function $\chi_V(r)$, whose Fourier transform is the spectral density $\tilde{\chi}_V(k)$, is shown in~\autoref{fig:autocorr}.
In 2D and for small $r$, the autocorrelation function can be expanded as~\cite{To02a}
\begin{equation}
\label{eq:autocorr_approx}
    \chi_V(r) = \phi_1 \phi_2 - rs/\pi + O(r^2),
\end{equation}
where $\phi_i$ is the volume fraction of phase $i$, and $s$ is the specific surface separating the two phases. As shown in~\autoref{fig:autocorr}, the small-$r$ approximation (dashed colored lines) captures the behavior of $\chi_V(r)$ at short distances. 
The different values of $\chi_V(r=0)$ are a consequence of different grain sizes; in particular, media derived from LS packings with $\epsilon=0.001$ have, on average, larger grains (see~\autoref{sec:grain_size}). 
The differences in specific surface induced by the different packing fractions are also reflected in the ordering of the spectral-density curves at intermediate wave numbers.

Importantly, although these two-phase media are not hyperuniform, even when generated from SHU packings, they still display long-range correlations, as indicated by oscillations of $\chi_V(r)$ at large $r$. This contrasts with media such as the random checkerboard, whose autocorrelation function is shown by the red dashed curve (see Ref.~\cite{To02a} for the analytic expression). For the random checkerboard and related random two-phase media, the autocorrelation function vanishes at $O(1)$ distances from the origin and remains non-negative for all $r$.

\subsection{Grain-size distribution}
\label{sec:grain_size}

\begin{figure}
    \centering
    \includegraphics[width=\linewidth]{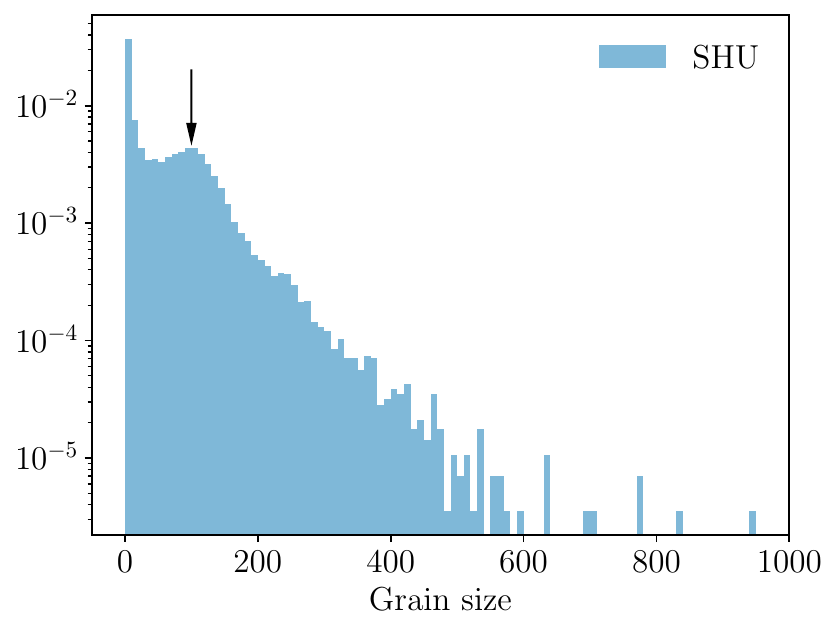}
    \includegraphics[width=\linewidth]{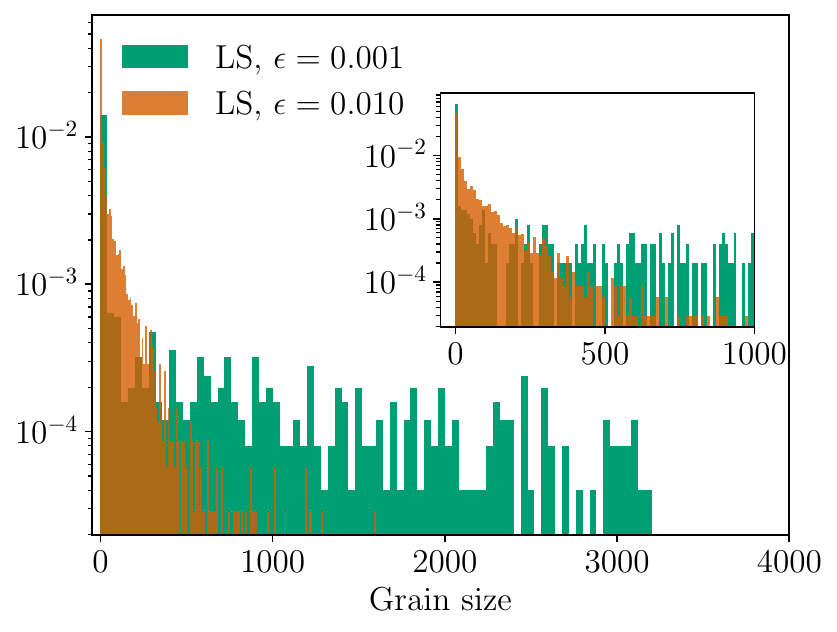}
    \caption{Average grain-size (number of disks in each grain) distribution of polycrystals derived from disk packings with $\phi=0.86$. (Top) SHU polycrystals display a characteristic grain size, revealed by a local maximum in the distribution (arrow). (Bottom) LS polycrystals do not show such a local maximum. LS polycrystals with $\epsilon=0.001$ have substantially larger grains than the other cases shown. The inset shows the distribution on the same scale as the SHU case, and makes visible the absence of a local maximum in the distribution.}
    \label{fig:average_size}
\end{figure}

We now compare the grain-size distributions of SHU and LS polycrystals. For each realization, we compute the number of disks in each grain, giving the grain size, and build the distribution across samples. The results are shown in~\autoref{fig:average_size}.

The two panels of~\autoref{fig:average_size} reveal a qualitative difference between the two classes of polycrystals. In the SHU case (top), the distribution is nonmonotonic and develops a clear local maximum at an intermediate grain size, signaling the emergence of a characteristic grain scale. By contrast, LS polycrystals (bottom) display broad, monotonic-like distributions without a corresponding peak, for both $\epsilon = 0.01$ and $\epsilon=0.001$. Because SHU configurations suppress long-wavelength density fluctuations by construction, grain nucleation and growth are not independent over large distances but are constrained by long-range correlations. These correlations disfavor both very small and very large grains and statistically select an intermediate size. Nonhyperuniform LS packings lack this global constraint and therefore do not exhibit a sharply selected grain scale.

In addition, we observe that slower compression rates in the LS algorithm lead to substantially larger grains. Physically, reducing the growth rate gives the system more time to relax through collisions, anneal local defects, and reorganize into more coherent crystalline domains before jamming. This trend is consistent with the well-known limit in which approaching a perfect lattice requires extremely slow compression.

A direct comparison with absolute grain sizes in experimental 2D polycrystals is not meaningful here because the disk diameter is the only microscopic length and neither protocol represents a material-specific coarsening time. We therefore compare the distributions at fixed $N$ and $\phi$ and focus on their shape, rather than assigning a physical size to the grains.

\section{Physical properties}
\label{sec:physical}

In this section, we examine how microstructure affects the physical properties of SHU polycrystals and compare their behavior with that of LS polycrystals. We focus on two quantities of direct relevance to materials applications: the time-dependent diffusion spreadability, which characterizes transport properties and is discussed in~\autoref{sec:diff_spread}, and the effective dynamic dielectric constant, which characterizes optical response and is discussed in~\autoref{sec:optical}.

\subsection{Diffusion spreadability}
\label{sec:diff_spread}

Consider a two-phase medium in which a solute is initially confined to phase 2 and subsequently diffuses into the surrounding liquid phase (phase 1). The time-dependent diffusion spreadability quantifies how efficiently the underlying microstructure facilitates this process.
For simplicity, we assume that the diffusion coefficient $D$ is identical in both phases at all times. This assumption can be relaxed to unequal diffusion coefficients without altering the long-time behavior; see Ref.~\cite{To21d}.

Formally, the time-dependent spreadability $\mathcal{S}(t)$ is defined as the fraction of total solute that has diffused into phase 1 at time $t$. A larger value of $\mathcal{S}(t)$ therefore indicates faster solute spreading. By definition, the asymptotic value is $\mathcal{S}(\infty)=\phi_1$. Torquato~\cite{To21d} showed that the excess spreadability in a two-phase medium in $d$-dimensional Euclidean space $\mathbb{R}^d$ can be expressed exactly as
\begin{equation}
\label{eq:exc_spread}
    \mathcal{S}(\infty) - \mathcal{S}(t) = \frac{1}{(2\pi)^d \phi_2} \int_{\mathbb{R}^d} \tilde{\chi}_V(\mathbf{k}) e^{-k^2 D t}d \mathbf{k}.
\end{equation}
This expression depends only on the one-point ($\phi_i$) and two-point [$\tilde{\chi}_V(\mathbf{k})$] structural correlations.

If the spectral density behaves as $\tilde{\chi}_V(\mathbf{k})\sim B|\mathbf{k}|^{\alpha}$ in the limit $|\mathbf{k}|\to 0$, the long-time decay of the excess spreadability obeys~\cite{To21d,yuan2026spreadability}
\begin{align}
    \mathcal{S}(\infty) - \mathcal{S}(t) =& \frac{C}{(Dt/a^2)^{(d+\alpha)/2}}\notag\\
    &+ o\left( (Dt/a^2)^{-(d+\alpha)/2} \right),\quad Dt/a^2 \gg 1,\label{eq:spread_general}
\end{align}
with
\begin{equation}
    C= B \ \Gamma \left(\frac{d+\alpha}{2} \right) \frac{1}{2^d \phi_2 \pi^{d/2} \Gamma(d/2)}.
\end{equation}
Thus, for fixed spatial dimension $d$, larger values of $\alpha$ lead to a faster approach of $\mathcal{S}(t)$ to its asymptotic limit $\phi_1$. In stealthy systems, including polycrystalline SHU packings, where effectively $\alpha=\infty$, the approach becomes exponential~\cite{To21d}:
\begin{equation}
\label{eq:spread_stealthy}
    \mathcal{S}(\infty) - \mathcal{S}(t) \sim \left[\frac{d v_1(1)}{2(2\pi)^d} \tilde{\alpha}_2(K;a) S(K)K^d\right] \frac{e^{-K^2 Dt}}{K^2 Dt},
\end{equation}
for $Dt/a^2 \gg 1$.

Note that spreadability is directly connected to experimental observables in nuclear magnetic resonance (NMR) and diffusion magnetic resonance imaging (dMRI). In particular, Torquato~\cite{To21d} showed that the pulsed-field-gradient spin-echo (PFGSE) amplitude $\mathcal{M}(\mathbf{k},t)$ of a fluid-saturated porous medium~\cite{Mit92b,Mi93,Or02} can be related to Eq.~\eqref{eq:exc_spread} through the mapping $\mathcal{S}(\infty) - \mathcal{S}(t) \to \mathcal{M}(\mathbf{0},t) - \phi_2$. Similarly, the transverse relaxation rate $dR_2/dt$ exhibits long-time asymptotic behavior analogous to excess spreadability~\cite{Ruh_2023_Observation}.
In the context of dMRI, it has been argued~\cite{No14} and experimentally confirmed~\cite{Le20} that $\mathcal{D}(t) - \mathcal{D}_e \sim C\,t^{-(d+\alpha)/2}$, where $\mathcal{D}(t)$ is the effective time-dependent diffusion coefficient, which coincides with the long-time scaling of $\mathcal{S}(\infty)-\mathcal{S}(t)$. This correspondence suggests the mapping $\mathcal{S}(t)\to\mathcal{D}(t)$ and $\mathcal{S}(\infty)\to\mathcal{D}_e=\mathcal{D}(\infty)$, highlighting the direct experimental relevance of diffusion spreadability for NMR and dMRI studies of physical and biological porous media.

\begin{figure}
    \centering
    \includegraphics[width=\linewidth]{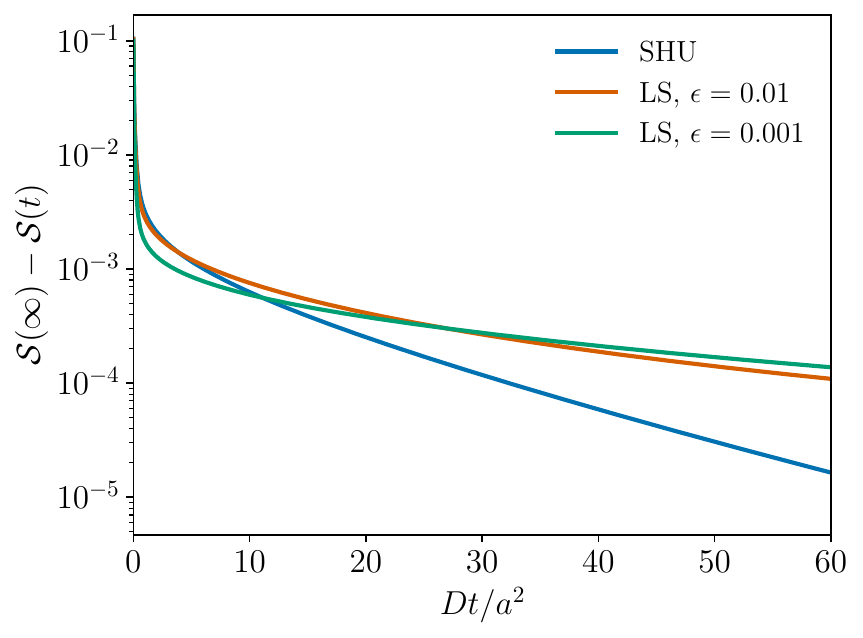}
    \caption{Excess diffusion spreadability for 2D SHU and LS disk packings with $\phi = 0.86$. For LS packings, the excess spreadability decays similarly for $\epsilon=0.01$ and $\epsilon=0.001$, consistent with Eq.~\eqref{eq:spread_general} with $\alpha=0$. The SHU case displays exponential decay, in agreement with Eq.~\eqref{eq:spread_stealthy}, indicating faster solute spreading from disks (phase 2) to matrix (phase 1).}
    \label{fig:diffspread}
\end{figure}

In Fig.~\ref{fig:diffspread}, we show the excess diffusion spreadability $\mathcal{S}(\infty) - \mathcal{S}(t)$ as a function of dimensionless time $Dt/a^2$ for two-phase media consisting of 2D disk packings embedded in a matrix. We report results for SHU and LS packings (with $\epsilon = 0.01$ and $\epsilon=0.001$) at packing fraction $\phi=0.86$.
While the short-time behavior is similar across systems, the decay for LS polycrystals becomes considerably slower than for SHU polycrystals at long times. This is a direct consequence of stealthy hyperuniformity: the decay is algebraic ($\sim t^{-1}$) for LS polycrystals, as in Eq.~\eqref{eq:spread_general}, but exponential for SHU polycrystals, according to Eq.~\eqref{eq:spread_stealthy}.
Overall, these results indicate that SHU polycrystals support more efficient long-time transport than conventional polycrystals at the same packing fraction. In this sense, stealthy hyperuniformity offers a practical route to improved transport performance in polycrystalline media.

\subsection{Optical transparency}
\label{sec:optical}

In this subsection, we quantify the optical properties of SHU polycrystals treated as a two-phase medium composed of disks and the surrounding matrix. The results presented here are based on the second-order approximation of the nonlocal strong-contrast expansion that is reviewed in Sec.~\ref{sec:strongcontrast_background}. 
\begin{figure}
    \centering
    \includegraphics[width=\linewidth]{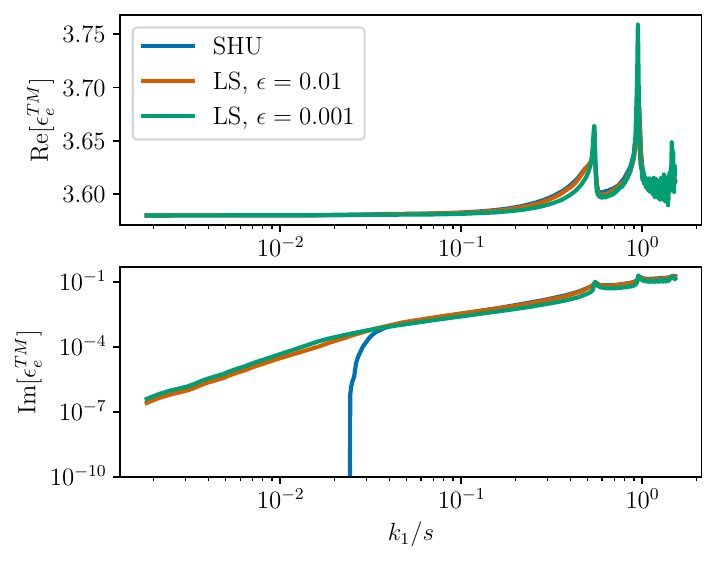}
    \includegraphics[width=\linewidth]{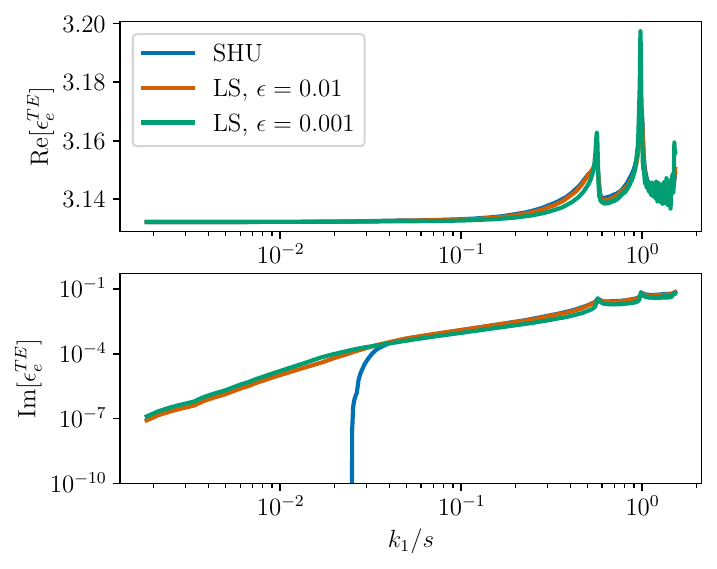}
    \caption{Effective dynamic dielectric constant for transversely isotropic media derived from SHU and LS disk packings with contrast ratio $\epsilon_2/\epsilon_1 = 4$ and $\phi = 0.86$. The top panel shows the real and imaginary parts of the transverse magnetic (TM) polarization, and the bottom panel shows the real and imaginary parts of the transverse electric (TE) polarization. The wave number on the horizontal axis is rescaled by the specific surface $s$ to make it dimensionless. For SHU media, the imaginary parts of both TE and TM responses vanish at small wave numbers, signaling perfect transparency over the corresponding wavelength range.}
    \label{fig:EDDC}
\end{figure}
In Fig.~\ref{fig:EDDC}, we show the TE and TM effective dynamic dielectric responses for a 3D transversely isotropic medium derived from 2D polycrystals. We obtain the TM and TE responses by numerically evaluating the second-order nonlocal strong-contrast expressions in Eqs.~\eqref{eq:TM_2D} and~\eqref{eq:TE_2D}, with the second-order coefficients computed from the spectral density. Focusing on the imaginary parts (lower panels of Fig.~\ref{fig:EDDC}), we observe that they vanish exactly over a wave number interval $0\leq k_1 < K_T$ for SHU polycrystals. This implies perfect transparency over the corresponding range of wavelengths~\cite{To21a,Kim_2024_extraordinary,Vanoni2025Dynamical}, with $K_T$ given by
\begin{equation}
    \frac{K_T}{K} = \frac{1}{2 \sqrt{\epsilon_*/\epsilon_q}},
\end{equation}
and $\epsilon_* = \langle \epsilon \rangle$ for TM polarization and $\epsilon_* = \epsilon_{BG}^{(2D)}$ for TE polarization.

This behavior is particularly remarkable when contrasted with conventional disordered or polycrystalline media, in which structural fluctuations generically produce nonzero attenuation and thus a finite imaginary part of the effective dielectric response at long wavelengths. In SHU polycrystals, stealthy constraints suppress the relevant long-wavelength scattering channels, thereby preserving transparency despite grain boundaries. Consequently, these materials combine polycrystalline morphology with crystal-like wave propagation, yielding reduced optical loss and improved transmission in the low-$k$ regime.

\section{Discussion and Conclusion}
\label{sec:conclusion}

In this work, we characterized a previously unexplored class of two-dimensional disk packings, namely, ultradense stealthy hyperuniform (SHU) polycrystalline arrangements of identical hard disks, introduced in Ref.~\cite{Kim_2025_DenseSphere}. These configurations were generated numerically via collective-coordinate optimization at small but finite stealthiness, $\chi=0.0025$, and were systematically benchmarked against polycrystals produced via the modified Lubachevsky--Stillinger (LS) jamming algorithm. This side-by-side comparison shows how stealthy hyperuniformity is reflected in the distribution of the size of grains in polycrystals, and leads to enhanced diffusive transport and optical properties compared to typical polycrystals.

Our results demonstrate that SHU packings can be genuinely polycrystalline, with grains and grain boundaries, while preserving the defining stealthy hyperuniform feature of vanishing structure factor over a finite neighborhood of the origin. In reciprocal space, this distinction appears directly in the low-$k$ regime of $S(k)$; in real-space morphology, it is reflected in grain statistics and intergrain organization. For example, the SHU samples display a characteristic grain-size scale associated with long-range intergrain correlations. In contrast, LS packings exhibit broader, nonselected grain-size distributions consistent with nonhyperuniform behavior.

Beyond structural characterization, we find that the stealthy constraints lead to distinct advantages in physical response. For diffusion, SHU polycrystals exhibit an exponential long-time decay of excess spreadability, while LS polycrystals exhibit the slower algebraic decay expected for nonhyperuniform media. This difference implies enhanced long-time transport efficiency in SHU microstructures at the same packing fraction. For wave transport, the second-order nonlocal strong-contrast formalism predicts a finite low-wave number optical transparency window in both TM and TE polarizations for SHU-derived media, where the imaginary part of the effective dynamic dielectric response vanishes.  By contrast, LS-derived media retain finite attenuation in the same regime. 

Taken together, these findings establish SHU polycrystalline packings as a distinct class of microstructures. They are neither conventional polycrystals, whose randomly distributed grains typically generate long-wavelength density fluctuations, nor perfect crystals, whose hyperuniformity follows from global periodic order. Instead, they combine local crystalline grains and grain-boundary defects with stealthy hyperuniform suppression of large-scale density fluctuations. Thus, we see that hyperuniformity can survive in explicitly defective polycrystalline structures, rather than being restricted to perfect crystals or fully amorphous disordered hyperuniform states.

The associated transport and optical calculations further show that this geometric organization has functional consequences, including faster long-time diffusion spreadability and a finite low-wave number transparency window relative to nonhyperuniform polycrystalline packings. Although the systems studied here are idealized disk-packing models, they suggest a route toward defect-tolerant multifunctional media in which grain boundaries need not destroy the advantages associated with hyperuniform order. Future work could explore experimentally realizable two-phase architectures, including additively manufactured metamaterials~\cite{ASKARI2020101562,Sh15,tumbleston_continuous_2015}, as well as probe the existence and properties of multicomponent or polydisperse ultradense SHU polycrystalline packings.

Several further comparisons follow naturally from the present structures. Hyperuniform polycrystalline arrays could be tested as substrates for frictional transport or vortex pinning, extending studies of random hyperuniform pinning arrays~\cite{Thien17}, or as porous media whose permeability is compared with that of other hyperuniform media~\cite{To20}. Their yield, fracture, and melting behavior may also differ from that of LS polycrystals and, in general, common polycrystals, but the present geometric model does not determine the direction of these changes. Pair-potential or machine-learning-assisted inverse design could be used to search for interactions that self-assemble the target structures~\cite{Adorf2018,Carleo2019}; the constrained objective in Eq.~\eqref{eq:soft_core} does not by itself establish such spontaneous assembly.

\acknowledgments

The authors are grateful to Samuel Dawley and  Murray Skolnick for helpful discussions and comments.
This work was supported by the Army Research Office under Cooperative Agreement No. W911NF-22-2-0103.

\section*{Data availability}

There are no publicly available research data or software supporting this manuscript. Requests for further information or data should be sent to the authors.

\appendix

\section{2D Structure factor}
\label{app:sec:2d_FT}

\begin{figure*}[!t]
    \centering
    \includegraphics[width=0.3\linewidth]{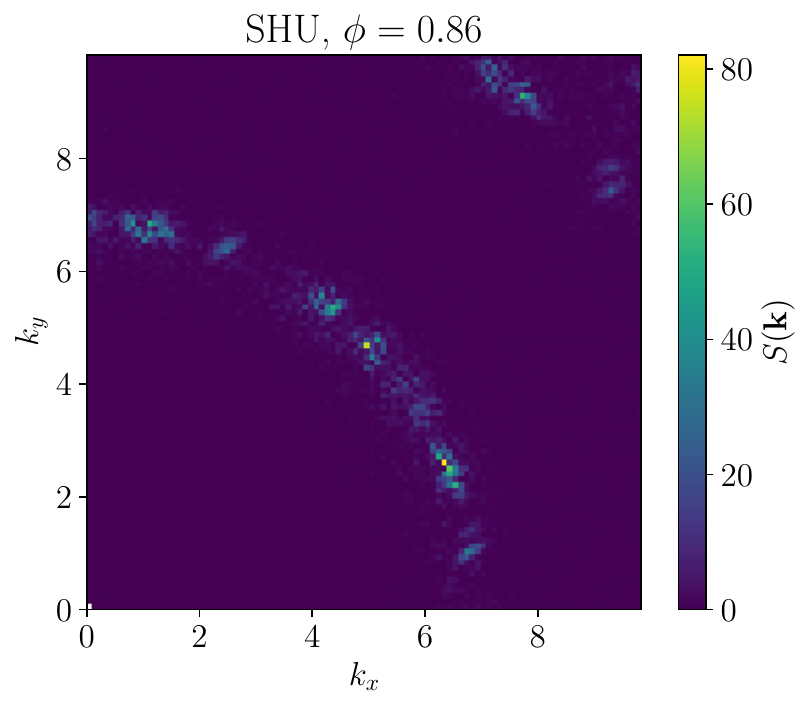}
    \includegraphics[width=0.3\linewidth]{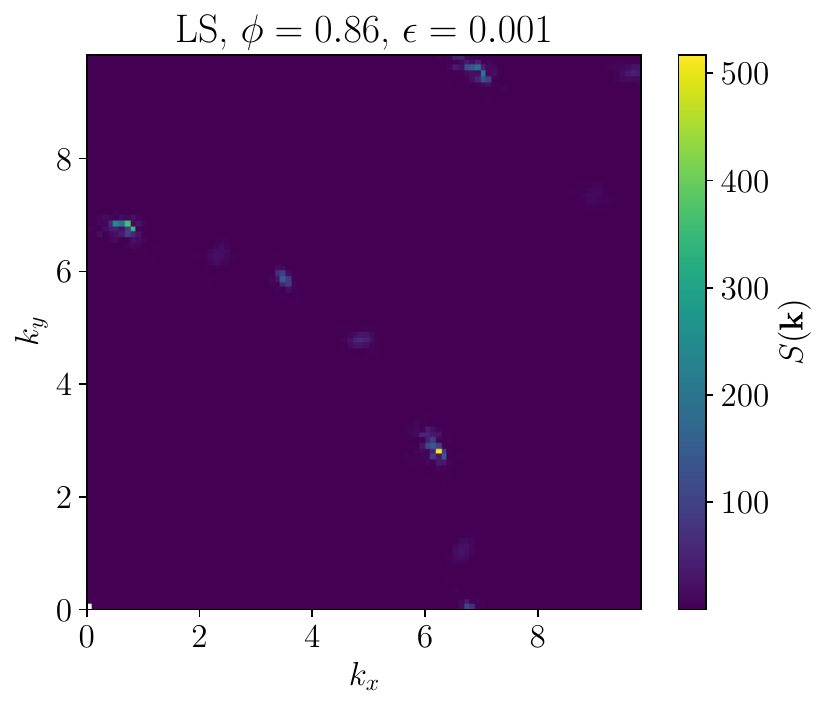}
    \includegraphics[width=0.3\linewidth]{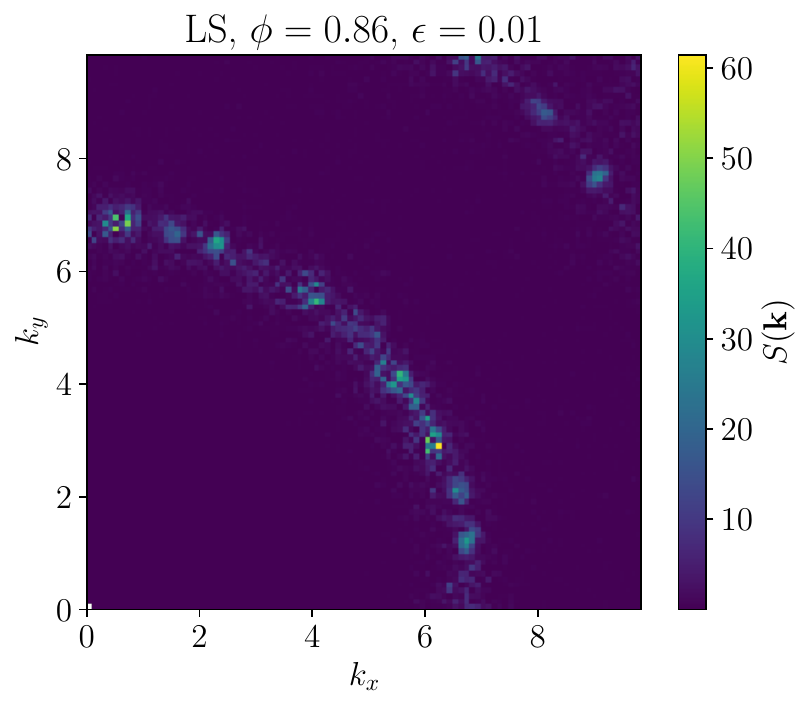}
    \includegraphics[width=0.3\linewidth]{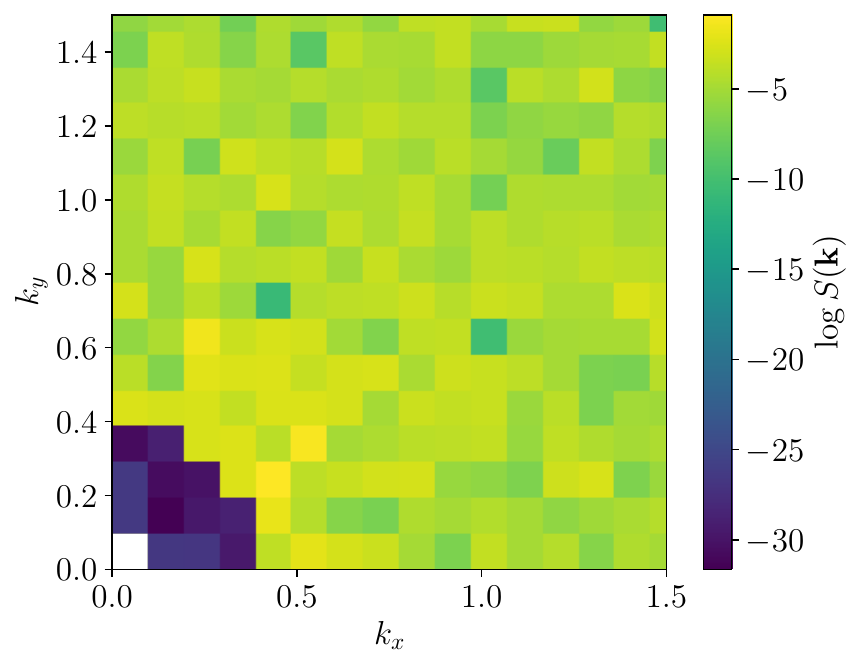}
    \includegraphics[width=0.3\linewidth]{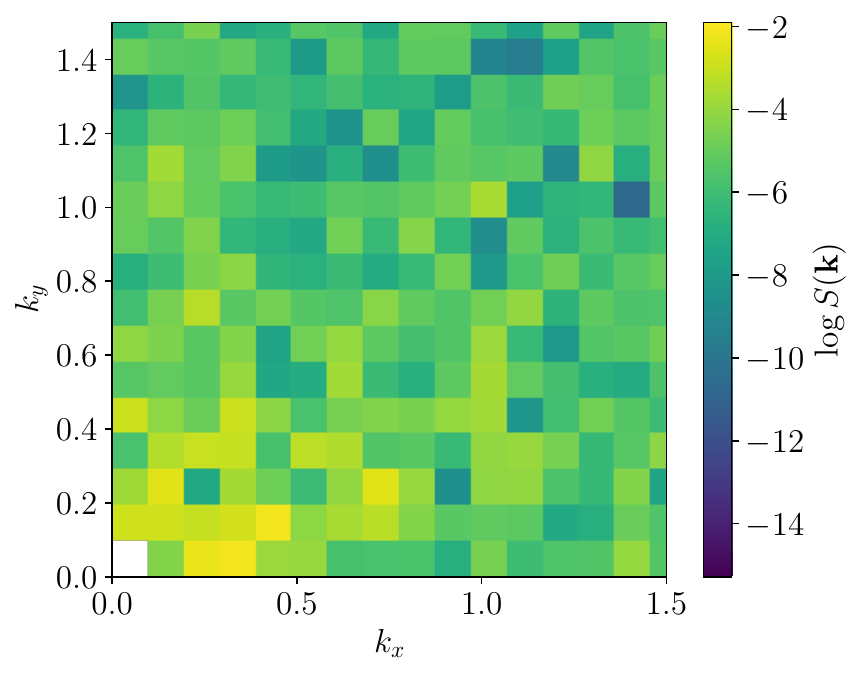}
    \includegraphics[width=0.3\linewidth]{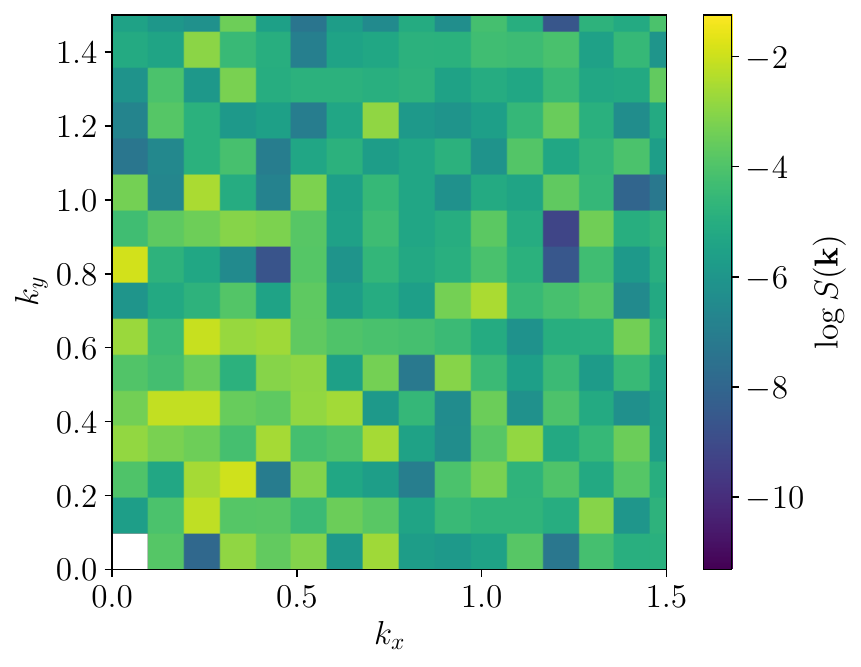}
    \caption{Plot of the two-dimensional structure factor of the polycrystals studied in the main text. Different columns refer to different models (SHU, LS with $\epsilon=0.001$, and LS with $\epsilon=0.01$), while the bottom row shows a zoom in closer to the origin compared to the top row. From the first row, we observe that the crystalline Bragg peaks are smeared due to variations in grain size and orientation, as well as grain boundaries. However, the LS polycrystals with $\epsilon=0.001$ retain a much more pronounced peak, as a consequence of the larger grain sizes, as shown in Fig.~\ref{fig:average_size}. The bottom row instead shows how the structure factor vanishes around the origin in SHU polycrystals, as prescribed by definition.}
    \label{fig:structfact2D}
\end{figure*}

In this Appendix, we provide the full two-dimensional structure factor for the representative polycrystals analyzed in the main text at packing fraction $\phi=0.86$. The three columns of Fig.~\ref{fig:structfact2D} correspond to SHU packings, LS packings with $\epsilon=0.001$, and LS packings with $\epsilon=0.01$, respectively. 

The first (upper) row shows a broad reciprocal-space view. In all three cases, the ideal sharp Bragg signatures of a perfect crystal are replaced by broadened and partially smeared maxima, reflecting the finite grain-size distribution, grain misorientation, and the presence of grain boundaries. The broadening is not uniform across models: the LS sample generated at the slower compression rate ($\epsilon=0.001$) exhibits sharper and more intense peaks than the LS sample at $\epsilon=0.01$, consistent with the larger average grain size discussed in Sec.~\ref{sec:grain_size}.

The second (lower) row zooms into the neighborhood of the origin to emphasize long-wavelength behavior, which is the most relevant regime for hyperuniformity. In the SHU case, the structure factor is identically zero over a finite region around $\mathbf{k}=\mathbf{0}$, revealing a clear stealthy exclusion zone. By contrast, both LS cases retain finite spectral weight at small wave numbers, indicating that long-wavelength density fluctuations are not suppressed to the same extent and confirming their nonhyperuniform character.

\section{Representative polycrystalline configurations}
\label{app:configurations}

\begin{figure*}[!t]
    \centering
    \includegraphics[width=0.32\linewidth]{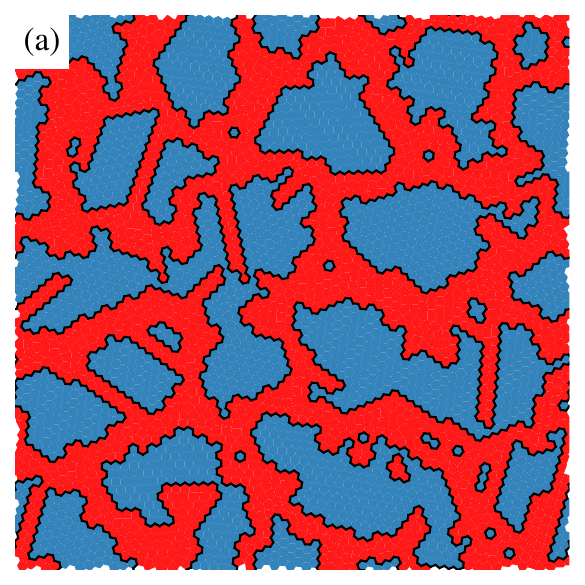}
    \includegraphics[width=0.32\linewidth]{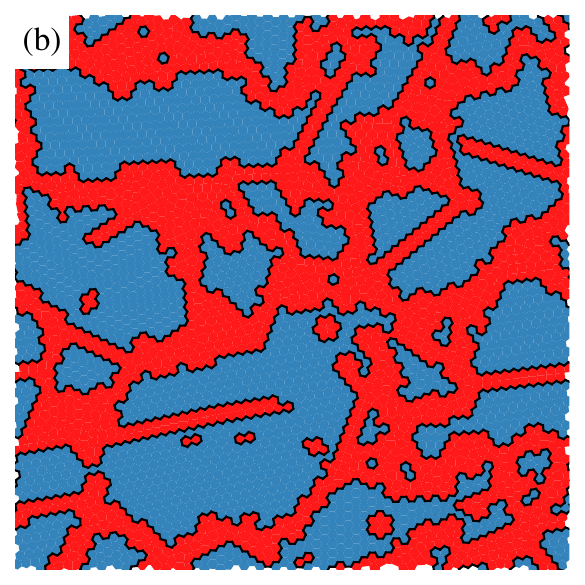}
    \includegraphics[width=0.32\linewidth]{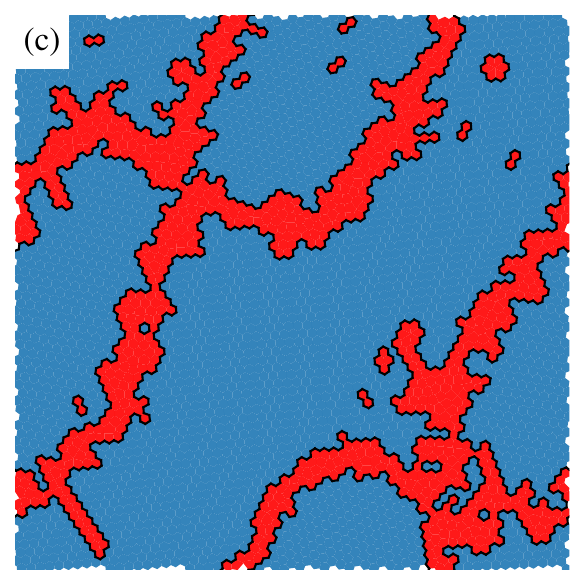}
    \caption{Representative polycrystalline configurations at $\phi=0.86$: (a) SHU with $\chi=0.0025$, (b) LS with $\epsilon=0.01$, and (c) LS with $\epsilon=0.001$. Blue disks belong to grains according to the criterion in Sec.~\ref{sec:grain_membership}, while red disks belong to grain boundaries. The slower LS compression in panel (c) produces visibly larger grains than the other two protocols.}
    \label{fig:representative_configurations}
\end{figure*}

Figure~\ref{fig:representative_configurations} displays the three types of structures used throughout the comparisons in the main text. The SHU and faster-compressed LS samples have comparable local grain scales, although their long-wavelength density fluctuations differ, whereas the slower-compressed LS sample contains much larger crystalline domains.

\section{Grain-orientation distributions}
\label{app:grain_orientations}

\begin{figure*}[!t]
    \centering
    \includegraphics[width=0.92\linewidth]{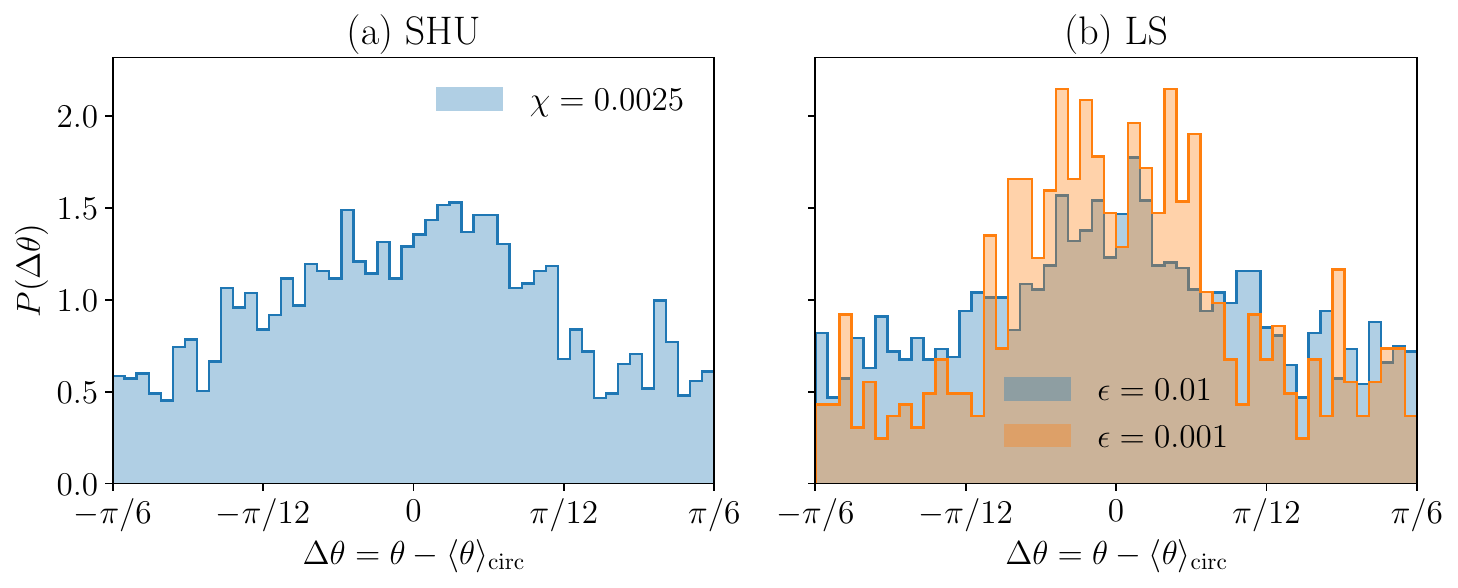}
    \caption{Probability density of the grain-orientation deviation $\Delta\theta$ for $N=4000$ and $\phi=0.86$: (a) SHU with $\chi=0.0025$ and (b) LS with $\epsilon=0.01$ and $\epsilon=0.001$. For each grain, the orientation is reduced modulo $\pi/3$. Within each realization, we subtract the sixfold circular mean and wrap the difference to $[-\pi/6,\pi/6)$. The distributions are pooled over 100 configurations and contain 3591, 3256, and 778 grains, respectively.}
    \label{fig:grain_orientation_distribution}
\end{figure*}

To compare the spread of grain orientations, we estimate one representative orientation per grain from a nearest-neighbor bond direction and report grain orientations relative to the sixfold circular mean. Figure~\ref{fig:grain_orientation_distribution} shows that the SHU and LS $\epsilon=0.01$ distributions have comparable widths, whereas the LS $\epsilon=0.001$ distribution is more concentrated around its circular mean because the same finite-size system contains fewer, larger grains. The corresponding mean magnitudes of the sample-level sixfold orientation order parameter $|\langle\exp(6 i\theta)\rangle|$ are $0.220$, $0.189$, and $0.373$, respectively. This measurement characterizes the distribution of grain orientations within each ensemble; it does not measure the pairwise misorientation of grains that share a boundary and therefore does not classify twin, small-angle, or large-angle boundaries.

\end{document}